%% file: conf_paper.tex
\documentclass[11pt]{article}
\usepackage{graphicx}

\newcommand{\BABARPubYear}    {01}

\newcommand{\BABARConfNumber} {05}
\newcommand{\SLACPubNumber} {8929}

\input pubboard/babarsym

\def\Btohh     {\ensuremath{\B \to h^+h^{\prime -}}}

\def\Bztopizpiz     {\ensuremath{\Bz \to \piz\piz}}
\def\Bzbtopizpiz     {\ensuremath{\Bzb \to \piz\piz}}

\def\pipi     {\ensuremath{\pi\pi}}

\def\hh     {\ensuremath{h^+h^{\prime -}}}

\def\fpm {\ensuremath{f_{\pm}(\deltat)}}
\def\ilam {\ensuremath{{\cal I}m\lambda}}
\def\alam {\ensuremath{\left|\lambda\right|}}

\def\spipi {\ensuremath{S_{\pi\pi}}}
\def\cpipi {\ensuremath{C_{\pi\pi}}}
\def\de {\ensuremath{\Delta E}}

\def\diffD {\ensuremath{\Delta D}}

\def\Btag {\ensuremath{B_{\rm tag}}}
\def\Bhh {\ensuremath{B_{hh}}}
\def\Bflav {\ensuremath{B_{\rm flav}}}

\def\ttag {\ensuremath{t_{\rm tag}}}

 \long\def\inst#1{\par\nobreak\kern 4pt\nobreak
    {\it #1}\par\vskip 10pt plus 3pt minus 3pt}

\begin{document}
{\pagestyle{empty}

\begin{flushright}
\babar-CONF-\BABARPubYear/\BABARConfNumber \\
%\babar-PUB-\BABARPubYear/\BABARPubNumber \\
SLAC-PUB-\SLACPubNumber \\
%hep-ex/\LANLNumber \\
July, 2001 \\
\end{flushright}

\par\vskip 3cm

% Title of the paper
\begin{center}
\Large \bf Study of $\CP$-violating asymmetries in {\boldmath $B\to\pi^{\pm}\pi^{\mp},K^{\pm}\pi^{\mp}$} decays
\end{center}
\bigskip

\begin{center}
\large The \babar\ Collaboration\\
\mbox{ }\\
%\today
July 26, 2001
\end{center}
\bigskip \bigskip

% Abstract
\begin{center}
\large \bf Abstract
\end{center}
We present a preliminary measurement of the time-dependent $\CP$-violating
asymmetry parameters $\spipi$ and $\cpipi$ in neutral $B$ decays to the
$\pi^{\pm}\pi^{\mp}$ \CP\ eigenstate, and an updated preliminary measurement of the charge 
asymmetry ${\cal A}_{K\pi}$ in $\B\to K^{\pm}\pi^{\mp}$ decays.  
Event yields and $\CP$-violation parameters 
are determined simultaneously from a multidimensional unbinned maximum likelihood 
fit.  In a data sample consisting of approximately $33$ million $\FourS\to\BB$ decays 
collected with the \babar\ detector at the SLAC PEP-II asymmetric $B$ Factory, we
find $65^{+12}_{-11}$ $\pi^{\pm}\pi^{\mp}$ and $217\pm 18$ $K^{\pm}\pi^{\mp}$ candidates
and measure $\spipi = 0.03^{+0.53}_{-0.56}\pm 0.11$, 
$\cpipi = -0.25^{+0.45}_{-0.47}\pm 0.14$, and 
${\cal A}_{K\pi} = -0.07 \pm 0.08 \pm 0.02$, where the first error is statistical
and the second is systematic.

\vfill
\begin{center}
Submitted to the \\
$20^{\it th}$ International Symposium on Lepton and Photon Interactions at High Energies, \\
$7/23$--$7/28/2001$, Rome, Italy
\end{center}

\vspace{1.0cm}
\begin{center}
{\em Stanford Linear Accelerator Center, Stanford University, 
Stanford, CA 94309} \\ \vspace{0.1cm}\hrule\vspace{0.1cm}
Work supported in part by Department of Energy contract DE-AC03-76SF00515.
\end{center}
}

\newpage

% Input author list file
\input pubboard/authors_EPS2001

\section{Introduction}
\label{sec:Introduction}
In the Standard Model, all $\CP$-violating effects arise from a single
complex phase in the three-generation CKM quark-mixing matrix~\cite{CKM}.  
One of the central questions in particle physics is whether this
mechanism is sufficient to explain the pattern of \CP violation observed 
in nature.  Recent measurements of the parameter $\stwob$ by the
\babar\/~\cite{BabarSin2betaObs} and BELLE~\cite{BelleSin2betaObs} Collaborations 
establish that \CP\ symmetry is violated in the neutral $B$-meson 
system.  These measurements are in agreement with other direct 
measurements~\cite{alls2b}, as well as indirect constraints implied by 
measurements and theoretical estimates of the CKM matrix elements~\cite{allCKM}.  
In addition to measuring $\stwob$ more precisely, one of the primary goals of the 
$B$-Factory experiments in the future will be to measure the remaining angles 
($\alpha$ and $\gamma$) and sides of the Unitarity Triangle in order to further 
test whether the Standard Model description of \CP\ violation is correct.

The study of $B$ decays to charmless hadronic two-body final states
will play an increasingly important role in our understanding of \CP violation.  
In the Standard Model, the time-dependent $\CP$-violating asymmetry in
the reaction $B\to\pip\pim$ is related to the angle $\alpha$.  
In addition, observation of a significant asymmetry between the decay rates for 
$\Bz\to\Kp\pim$ and $\Bzb\to\Km\pip$ would be evidence for direct 
\CP violation, and ratios of branching fractions for various $\pi\pi$ 
and $K\pi$ decay modes are sensitive to the angle $\gamma$.  Finally,
branching fraction measurements provide critical tests of theoretical 
models that are needed to extract \CP\ information from the 
experimental observables.

The \babar\ Collaboration recently reported measurements of branching
fractions and charge asymmetries for several charmless two-body $B$ decays 
using a dataset of $22.6$ million $\BB$ pairs~\cite{twobodyPRL}.  
In this paper, using a data sample of approximately $33$ million 
$\BB$ pairs, we report preliminary measurements of the time-dependent 
$\CP$-violating asymmetry in neutral $B$ decays to the $\pip\pim$ 
\CP\ eigenstate, and the asymmetry between $\Bz\to\Kp\pim$ and 
$\Bzb\to\Km\pip$ decays.

\section{Data sample and \babar\ detector}
\label{sec:babar}

The data sample used in this analysis consists of $33.7\invfb$
collected with the \babar\ detector at the Stanford Linear Accelerator 
Center's PEP-II storage ring between October 1999 and June 2001.  
The PEP-II facility operates nominally at the \Y4S\ resonance, providing 
asymmetric collisions of $9.0\gev$ electrons on $3.1\gev$ positrons.  The 
dataset includes $30.4\invfb$ collected in this configuration (on-resonance) and 
$3.3\invfb$ collected below the \BB\ threshold (off-resonance) that are used for 
continuum background studies.  The on-resonance sample corresponds to approximately
33 million produced \BB\ pairs.

\babar\ is a $4\pi$ solenoidal spectrometer optimized for the asymmetric beam 
configuration and is described in detail elsewhere~\cite{ref:babar}.  
Charged particle (track) momenta are measured in a tracking system consisting of a 
5-layer, double-sided, silicon vertex tracker (SVT) and a 40-layer drift chamber (DCH)
filled with a gas mixture of helium and isobutane, both operating within a 
$1.5\,{\rm T}$ superconducting solenoidal magnet.  The typical decay vertex resolution
for fully reconstructed $B$ decays is approximately $65\mum$ along the 
center-of-mass (CM) boost direction.  Photons are detected in an 
electromagnetic calorimeter (EMC) consisting of 6580 CsI(Tl) crystals arranged 
in barrel and forward endcap subdetectors.  The iron flux return (IFR)
is segmented and instrumented with multiple layers of resistive plate chambers
for the identification of muons and long-lived neutral hadrons.

Tracks from the decay $B\to\hh$ are identified as pions or kaons by the 
Cherenkov angle $\theta_c$ measured by a detector of internally reflected 
Cherenkov light (DIRC).  The DIRC system is a unique type of Cherenkov detector 
that relies on total internal reflection within the radiator to deliver 
the Cherenkov light outside the tracking and magnetic volumes.  The typical 
separation between pions and kaons varies from $8\sigma$ at $2\gevc$ to 
$2.5\sigma$ at $4\gevc$, where $\sigma$ is the average resolution on $\theta_c$.  
Kaons used in $B$ tagging are identified with a combination of $\theta_c$ 
(for momenta down to $0.7\gevc$) and specific ionization ($dE/dx$) measurements in
the DCH and SVT.

\section{Analysis overview}
\label{sec:overview}

The time-dependent $\CP$-violating asymmetry in the decay $B\to\pip\pim$ 
arises from interference between mixing and decay amplitudes, and interference 
between the tree and penguin decay amplitudes. A $\Bz\Bzb$ pair produced in 
\FourS\ decay evolves in time in a coherent $P$-wave state until one of the two 
mesons decays.  We reconstruct the decay $\Btohh$ ($\Bhh$), where $h$ is a 
pion or kaon, and examine the remaining particles in the event to ``tag'' the 
flavor of the other $B$ meson (\Btag).  
%If \Btag\ decays to a $\Bz\,(\Bzb)$ at a 
%time \ttag, then \Bhh\ is known to be a $\Bzb\,(\Bz)$ at that same time.
Defining $\deltat = t_{hh} - \ttag$ as the time between the decays of \Bhh\
and \Btag, the decay rate distribution $f_+\,(f_-)$ when $\Bhh\to\pip\pim$
and \Btag\ is a $\Bz\,(\Bzb)$ is given by~\footnote{We assume $\Delta\Gamma = 0$.}
\begin{equation}
\fpm = \frac{e^{-\left|\deltat\right|/\tau}}{4\tau} \left[1 \pm \spipi\sin(\deltamd\deltat) \mp
\cpipi\cos(\deltamd\deltat)\right],
\label{fplusminus}
\end{equation}
where $\tau$ is the $\Bz$ lifetime, $\deltamd$ is the $\Bz\Bzb$ mixing frequency, 
and
\begin{equation}
\spipi = \frac{2\,\ilam}{1+\alam^2},\quad{\rm and,}\quad \cpipi = \frac{1-\alam^2}{1+\alam^2}.
\label{SandCdef}
\end{equation}
Ignoring the contribution from the penguin amplitude, the complex parameter $\lambda$ is
\begin{equation}
%\lambda \equiv \frac{q}{p}\frac{\bar{A}_{\pi\pi}}{A_{\pi\pi}} 
%= \eta_f e^{-2i\beta}\frac{\bar{A}_{\bar{f}}}{A_f},
%\label{lambdadef}
\lambda(B\to\pip\pim) \equiv \frac{q}{p}\frac{\bar{A}_{\pi\pi}}{A_{\pi\pi}} 
= \eta_{\pi\pi}\left(\frac{V_{tb}^*V_{td}}{V_{tb}V_{td}^*}\right)
\left(\frac{V_{ud}^*V_{ub}}{V_{ud}V_{ub}^*}\right),
\end{equation}
where $\eta_{\pi\pi} = +1$ is the \CP\ eigenvalue of the final state, and the assumption
of no \CP violation in mixing ($\left|q/p\right|=1$) is implicit.  Thus, in the absence
of penguins, $\alam = 1$ and $\ilam = \stwoa$, where 
$\alpha \equiv \arg\left[-V_{td}V_{tb}^*/V_{ud}V_{ub}^*\right]$.  
However, the $b\to d$ gluonic penguin amplitude 
carries the weak phase $\arg(V_{td}^*V_{tb})$ and, in general, modifies both $\alam$ and 
$\ilam$.  In this case, $\alam\ne 1$ and $\spipi$ becomes
\begin{equation}
\spipi = \frac{2\,\alam\sin{2\alpha_{\rm eff}}}{1+\alam^2},
\end{equation}
where $\alpha_{\rm eff}$ depends on the magnitudes and strong phases of the tree and
penguin amplitudes.  Recent theoretical estimates of the relative size of penguin 
and tree amplitudes vary~\cite{Beneke01a,PQCD}, but large effects are possible.

%where the first term arises from mixing and the second term involves the decay
%amplitdues $A_f\,(\bar{A}_f)$ is the amplitude for the decay 
%$\Bz\,(\Bzb)\to \pip\pim$, $\eta_f$ is 
%the \CP\ eigenvalue of the final state ($+1$), and the assumption of 
%no \CP\ violation in mixing ($\left|q/p\right|=1$) is implicit.  

%For decays involving $b\to u\bar{u}d$ transitions, the tree amplitude carries 
%the weak phase $\gamma = \arg(V_{ub}^*)$.  Ignoring the contribution from 
%the penguin amplitude and defining $\alpha$ + $\beta$ + $\gamma$ = $\pi$, $\lambda$ 
%can be written as
%\begin{equation}
%\lambda = e^{-2i(\beta + \gamma)} = e^{2i\alpha}.
%\label{alphadef}
%\end{equation}
%Thus, in the absence of penguins, the time-dependent $\CP$-violating asymmetry between
%$f_+$ and $f_-$ measures $\stwoa$.  However, the $b\to d$ gluonic penguin amplitude carries the 
%weak phase $-\beta = \arg(V_{td})$ and, in general, modifies both $\alam$ and $\ilam$.  In this 
%case, $\lambda$ becomes
%\begin{equation}
%\lambda \equiv \alam e^{2i\alphaeff},
%\end{equation}
%where $\alphaeff$ depends on the magnitudes and strong phases of the tree and penguin amplitudes.
%A recent calculation estimates $\left|P/T\right|\sim 0.3$~\cite{Beneke01a}, where 
%$P$ and $T$ are the penguin and tree amplitudes, respectively.

It is possible to extract $\alpha$ in the presence of penguins with 
little or no theoretical error using an isospin analysis~\cite{Gronau90a}
(see, however, Ref.~\cite{isospinbreaking}).  
The analysis requires measurements of the separate branching fractions for 
$\Bztopizpiz$ and $\Bzbtopizpiz$ as well as the charge-averaged branching 
fraction for $\B^{\pm}\to\pi^{\pm}\piz$.  However, it will be some time before such 
an analysis is experimentally feasible.  Alternatively, bounds on the 
penguin-induced shift in $\alpha$ can be derived from ratios of various 
two-body branching fractions~\cite{bounds}.
%without the need for tagging in 
%$\Bztopizpiz$.  
Finally, recent theoretical work allows the extraction of 
$\alpha$ given a measurement of $\spipi$~\cite{Beneke01a}.

In this analysis we extract signal and background yields for $\pi^{\pm}\pi^{\mp}$, 
$K^{\pm}\pi^{\mp}$, and $K^{\pm}K^{\mp}$ decays, and the amplitudes of the $\pi\pi$ sine 
($\spipi$) and cosine ($\cpipi$) oscillation terms simultaneously from an unbinned maximum 
likelihood fit.  We parameterize the $K\pi$ component in terms of the total yield and 
the $\CP$-violating charge asymmetry
\begin{equation}
{\cal A}_{K\pi} \equiv \frac{N_{\Km\pip} - N_{\Kp\pim}}{N_{\Km\pip} + N_{\Kp\pim}}.
\end{equation}
Including the more abundant $K\pi$ sample in the fit also allows for validation of the 
$\deltat$ parameterization from direct measurements of $\tau$ and $\deltamd$ 
(via mixing in $\Bz\to\Bzb\to\Km\pip$) in the same sample used to extract $\spipi$ and $\cpipi$.
In addition, background discrimination provided by the measurement of $\deltat$
improves the error on signal yields.  The combined fit to yields and \CP parameters therefore
facilitates the simultaneous optimization of branching fraction and \CP\ measurements, both
of which are necessary to extract reliable information about $\alpha$.

\section{Event selection}
\label{selection}

Hadronic events are selected based on track multiplicity and event
topology.  Tracks in the polar angle region $0.41 < \theta_{\rm lab} < 2.54$ with 
transverse momentum greater than $100\mevc$ are required to pass quality 
cuts, including number of drift chamber hits used in the track fit and impact 
parameter in the $r$--$\phi$ and $r$--$z$ planes, where the cylindrical coordinate 
$z$ is aligned along the detector axis in the electron beam direction.  At least 
three tracks must pass the above selection.  To reduce contamination from Bhabha 
and \mumu\ events the ratio of second to zeroth Fox-Wolfram 
moments~\cite{ref:foxwolf}, $R_2 = H_2/H_0$, is required to be less than $0.95$.  
Residual background from tau hadronic decays is reduced by requiring the 
sphericity~\cite{ref:sphericity} of the event to be greater than $0.01$.

Candidate $B\to\hh$ decays are reconstructed by combining pairs of oppositely-charged 
tracks (pion mass assumed) with a good quality vertex.  We require each track to
have an associated $\theta_c$ measurement with a minimum of six Cherenkov photons
above background.  Protons are rejected based on $\theta_c$ and electrons 
are rejected based on $dE/dx$, shower shape in the EMC, and the ratio of shower 
energy and track momentum.  Non-resonant $q\bar{q}$ background is suppressed by removing 
jet-like events from the sample: we define the CM angle $\theta_S$ between the sphericity 
axes of the $B$ candidate and the remaining tracks and photons in the event, and require 
$\left|\cos{\theta_S}\right|<0.8$, which removes approximately $83\%$ of the background.  
The total efficiency of the above selection on signal events is approximately $38\%$.

We define a beam-energy substituted mass 
$\mes = \sqrt{E^2_{\rm b}- {\mathbf {p}}_B^2}$.  The beam energy is defined in the
laboratory frame as $E_{\rm b} =(s/2 + {\mathbf {p}}_i\cdot {\mathbf {p}}_B)/E_i$, 
where $\sqrt{s}$ and $E_i$ are the total energies of the \epem\ system in the
CM and lab frames, respectively, and 
${\mathbf {p}}_i$ and ${\mathbf {p}}_B$ are the 
momentum vectors in the lab frame of the \epem\ system and the $B$ candidate, 
respectively.  Defining $\mes$ in the laboratory frame removes the dependence 
on the track mass hypothesis.  Signal events are Gaussian distributed in $\mes$ 
with a mean of $5.280\gevcc$ and a resolution of $2.6\mevcc$.  The background
shape is parameterized by a threshold function~\cite{ARGUS} with a fixed endpoint given by
the average beam energy.

We define a second kinematic variable $\de$ as the difference between the $B$ 
candidate energy in the CM frame and $\sqrt{s}/2$.  The $\de$ distribution is 
peaked near zero for $\pip\pim$ decays and shifted on average $-45\mev$ ($-91\mev$) 
for modes with one (two) charged kaons, where the exact separation depends on the 
laboratory kaon momentum.  The resolution on $\Delta E$ for signal decays is 
approximately $26\mev$.  The background is parameterized by a quadratic function.  

Candidate $B$ mesons selected in the region $5.2 < \mes < 5.3\gevcc$ 
and $\left|\de\right|<0.15\gev$ are used to extract yields and \CP parameters
from an unbinned maximum likelihood fit.  The total number of events in 
the fit region satisfying the above criteria is $9741$.  A sideband region, 
defined as $5.2 < \mes < 5.26\gevcc$ and $\left|\de\right|<0.42\gev$, is used to
extract various background parameters.

\section{Analysis}
\label{sec:analysis}

The analysis method combines the techniques used to measure charmless two-body
branching fractions~\cite{twobodyPRL} and the $\CP$-violating parameter 
$\stwob$~\cite{BabarSin2betaObs}.  
%A blind analysis technique was used, where the 
%signs and central values of $\spipi$ and $\cpipi$ were hidden until the selection
%criteria and dominant systematic errors were determined.  
The primary issues are
\begin{itemize}
 \item determining the flavor of the $\Btag$ meson;
 \item measuring the distance $\deltaz$ between the \Bhh\ and \Btag\ vertices;
%       The proper decay time diference $\deltat$ is obtained from the measurement of
%       $\deltaz$ and the known boost of the CM frame.
 \item discriminating signal from background;
 \item separating pions and kaons in the kinematically similar decays $\B\to\pi\pi,\, K\pi,\, KK$;
 \item extracting yields and \CP\ asymmetries with an unbinned maximum likelihood fit;
\end{itemize}
The first four issues have been described in previous publications.  In this section we
summarize the main points and describe the fit technique.

\subsection{Flavor tagging}
\label{sec:tagging}

We use the standard \babar\ $B$-tagging algorithm to determine the flavor of the \Btag\
meson~\cite{ref:babarmix}.  The algorithm relies on the correlation between the flavor of the 
$b$-quark and the charge of the remaining tracks in the event after removal of the 
\Bhh\ candidate.  Five mutually exclusive 
tagging categories are defined: {\tt Lepton}, {\tt Kaon}, {\tt NT1}, {\tt NT2}, and 
{\tt Untagged}.  {\tt Lepton} tags rely on primary 
electrons and muons from semileptonic $B$ decays, while {\tt Kaon} tags use the sum of the charges 
of all identified kaons.  The {\tt NT1} and {\tt NT2} categories are derived from a neural network 
that is sensitive to charge correlations between the parent \B\ and unidentified leptons and 
kaons, soft pions, or the charge and momentum of the track with the highest CM momentum. 
The addition of {\tt Untagged} events improves the signal yield estimates and provides a large 
sample for determinating background shape parameters directly in the maximum likelihood fit.

\begin{figure}[!tbp]
\begin{center}
\begin{minipage}[h]{8.0cm}
\includegraphics[width=8.0cm]{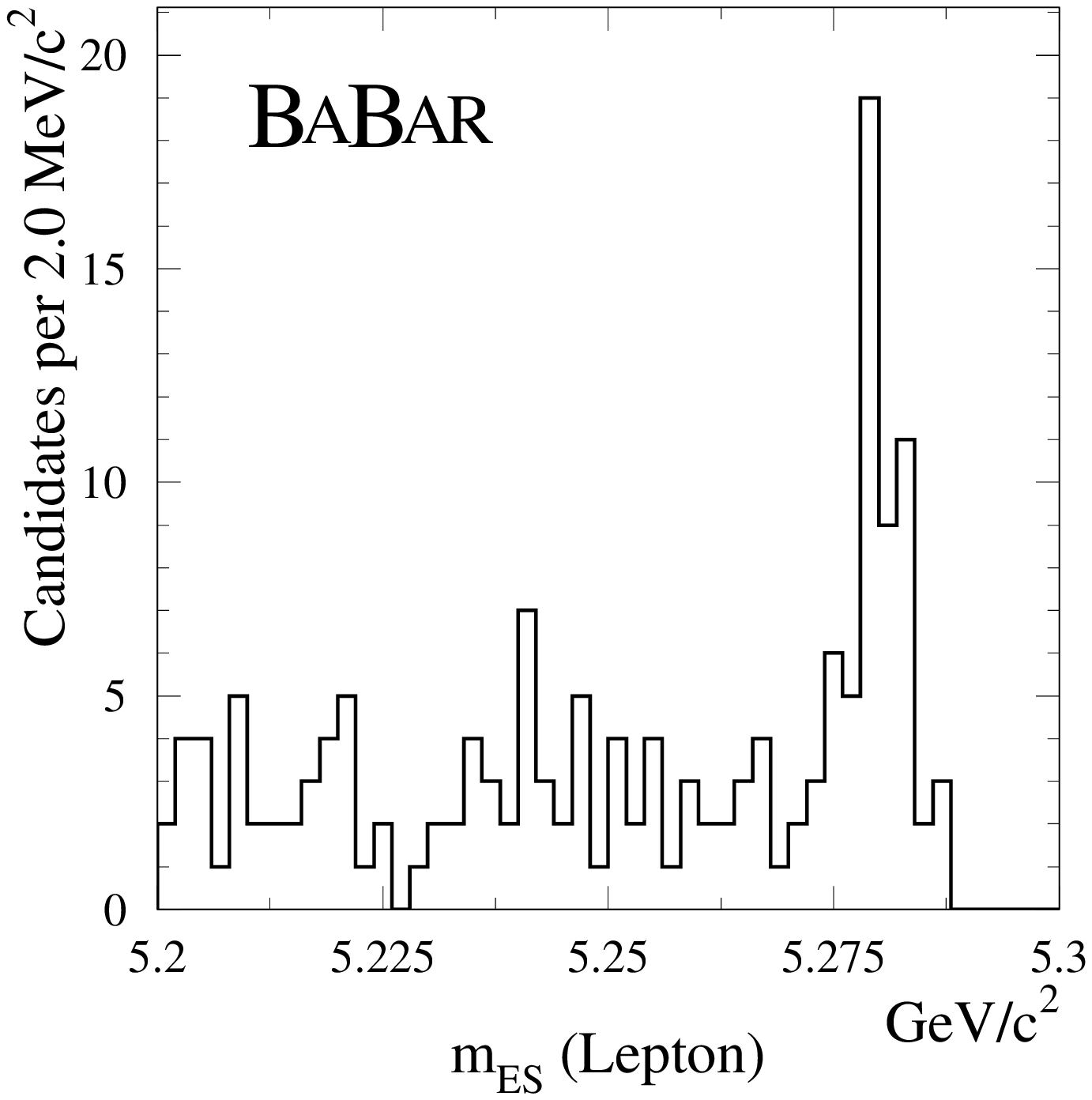}
\end{minipage}
\begin{minipage}[h]{8.0cm}
\includegraphics[width=8.0cm]{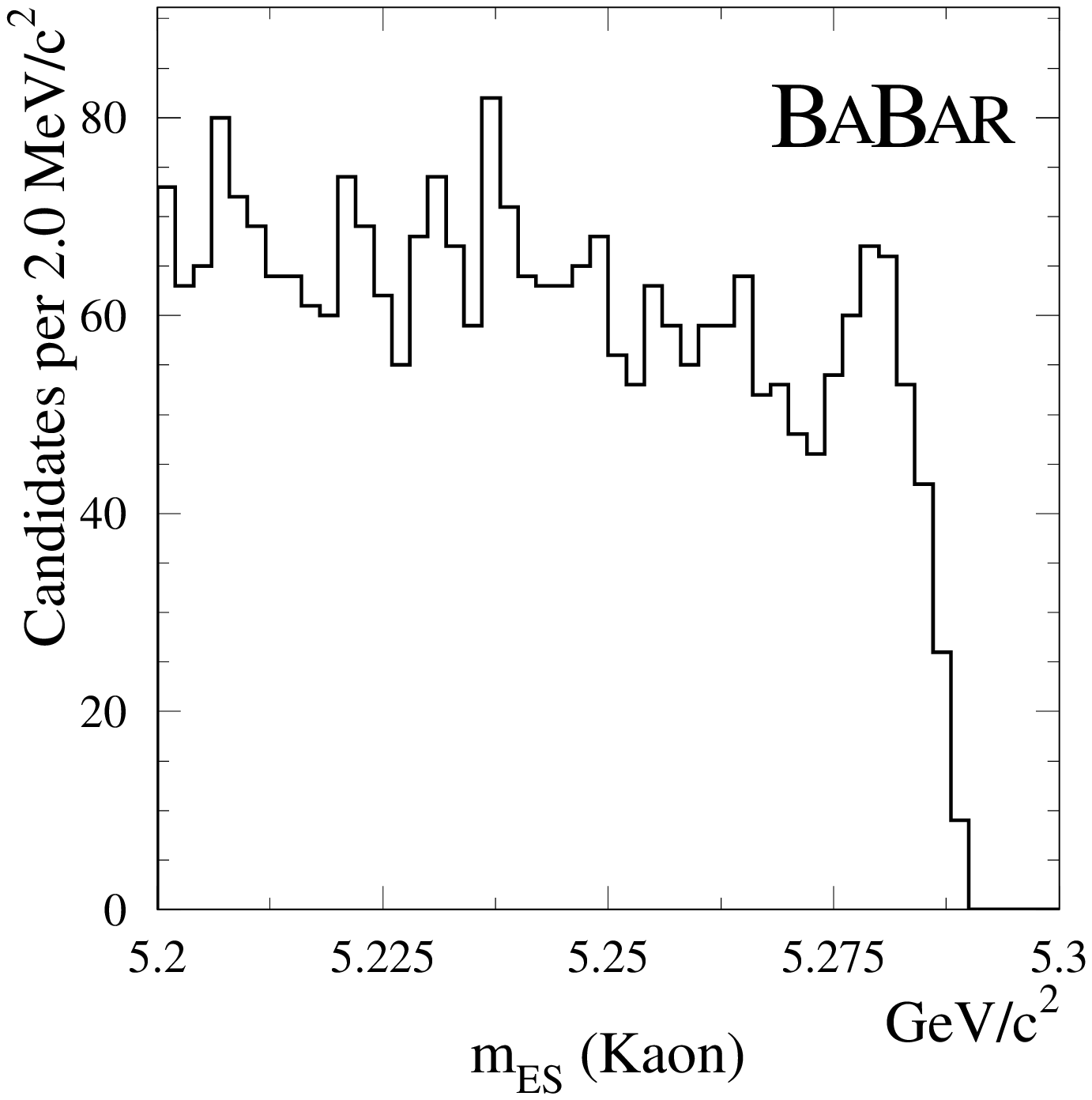}
\end{minipage}
\end{center}
\begin{center}
\begin{minipage}[h]{8.0cm}
\includegraphics[width=8.0cm]{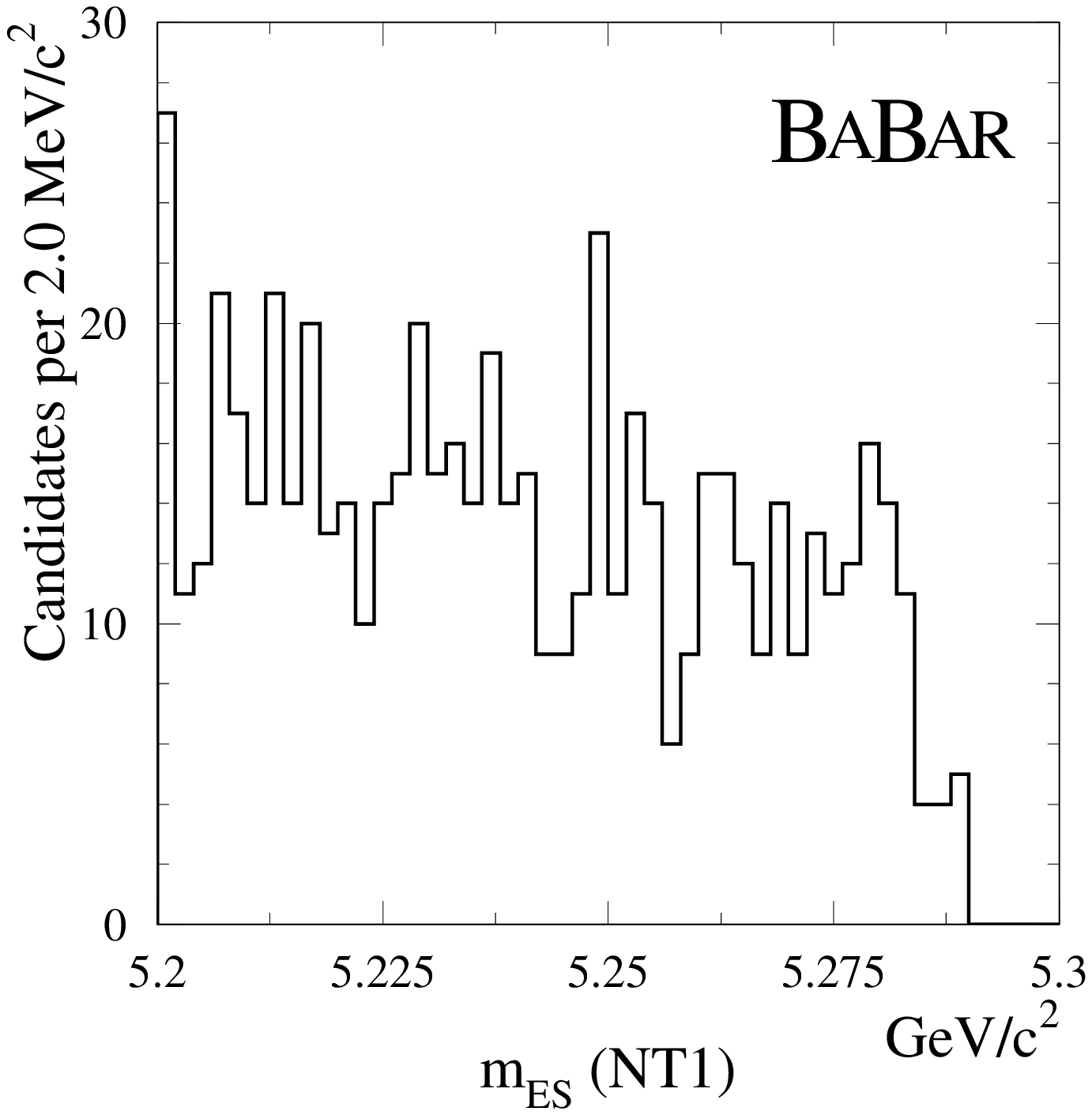}
\end{minipage}
\begin{minipage}[h]{8.0cm}
\includegraphics[width=8.0cm]{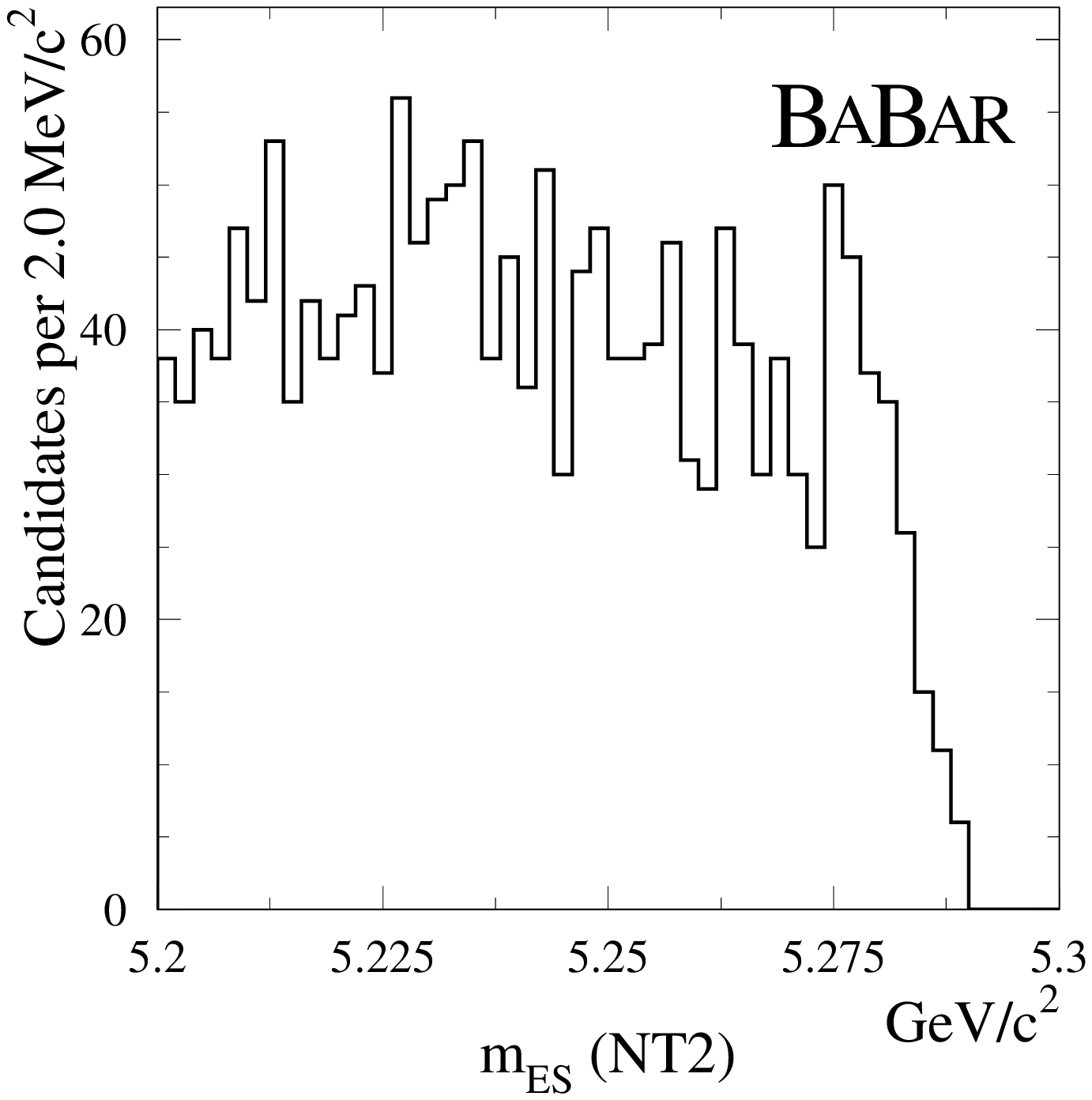}
\end{minipage}
\end{center}
\caption{Distributions of $\mes$ for $\hh$ events satisfying the
selection criteria and tagged in the {\tt Lepton}, {\tt Kaon}, {\tt NT1}, and
{\tt NT2} categories.}
\label{fig:nominalmes}
\end{figure}

The quality of tagging is expressed in terms of the effective efficiency 
$Q = \sum_i \epsilon_i D_i^2$, where $\epsilon_i$ is the fraction of events tagged in 
category $i$ and the dilution $D_i = 1-2w_i$ is related to the mistag fraction $w_i$.  
The statistical errors on $\spipi$ and $\cpipi$ are proportional to $1/\sqrt{Q}$.  Table~\ref{tab:tagging} 
summarizes the tagging performance in $\BB$ events, obtained from a sample 
\Bflav\ of fully reconstructed neutral $B$ decays into $D^{(*)-}h^+\,(h^+ = \pip, \rho^+, a_1^+)$ and
$\jpsi K^{*0}\,(K^{*0}\to\Kp\pim)$ flavor eigenstates~\cite{BabarSin2betaObs}.  We use the same 
tagging efficiencies and dilutions for signal $\pi\pi$, $K\pi$, and $KK$ decays.  Separate 
background tagging efficiencies for each species are obtained from a fit to the on-resonance 
sideband data and reported in Table~\ref{tab:bkgtag}.  The division of data into
tagging category and flavor is summarized in Table~\ref{tab:yields}, and the 
distributions of $\mes$ for events tagged in each category are shown in 
Fig.~\ref{fig:nominalmes}.

\begin{table}[!tbp]
\caption{Tagging efficiency, average dilution, dilution difference 
$\diffD = D(\Bz) - D(\Bzb)$, and effective tagging efficiency $Q$
for signal events in each tagging category, as measured in a sample
of neutral $B$ decays to flavor eigenstates.}
\smallskip
\begin{center}
\begin{tabular}{ccccc} \hline\hline
Category & $\epsilon\,(\%)$ & $D\,(\%)$ & $\diffD\,(\%)$ & $Q\,(\%)$ \rule[-2mm]{0mm}{6mm} \\\hline
{\tt Lepton}   & $11.0\pm 0.3$ & $82.3 \pm 2.7$ & $-2.1  \pm 4.5$ & $7.5\pm  0.5$ \rule[0mm]{0mm}{2mm}\\
{\tt Kaon}     & $35.8\pm 0.5$ & $64.8 \pm 2.0$ & $ 3.5  \pm 3.1$ & $15.0\pm 1.0$ \\
{\tt NT1}      & $8.0 \pm 0.3$ & $55.6 \pm 4.2$ & $-12.1 \pm 6.7$ & $2.5\pm  0.4$ \\
{\tt NT2}      & $13.9\pm 0.4$ & $30.2 \pm 3.8$ & $ 9.0  \pm 5.7$ & $1.3\pm  0.3$ \\
{\tt Untagged} & $31.3\pm 0.5$ & -- 	  & --            & --          \\ \hline
Total $Q$ & & & & $26.3\pm 1.2$ \rule[-2mm]{0mm}{6mm} \\\hline\hline
\end{tabular}
\end{center}
\label{tab:tagging}
\end{table}

\begin{table}[!tbp]
\caption{Tagging efficiencies $(\%)$ for background\,$(b)$ events in each species
$(\pi\pi, K\pi, KK)$ as determined from a fit to the on-resonance sideband data.}
\smallskip
\begin{center}
\begin{tabular}{cccccc} \hline\hline
Category & $\epsilon_b(\pi\pi)$ & $\epsilon_b(K\pi)$ & $\epsilon_b(KK)$ \rule[-2mm]{0mm}{6mm} \\\hline 
{\tt Lepton}   & $1.0\pm 0.1$  & $1.0\pm 0.1$  & $1.5\pm 0.2$  \\
{\tt Kaon}     & $26.0\pm 0.4$ & $33.1\pm 0.6$ & $23.5\pm 0.7$ \\
{\tt NT1}      & $6.6\pm 0.2$  & $5.4\pm 0.3$  & $6.9\pm 0.4$  \\
{\tt NT2}      & $17.6\pm 0.4$ & $15.3\pm 0.5$ & $19.7\pm 0.6$ \\
{\tt Untagged} & $48.9\pm 0.7$ & $45.2\pm 0.6$ & $48.3\pm 0.8$ \\\hline\hline
\end{tabular}
\end{center}
\label{tab:bkgtag}
\end{table}

\begin{table}[!tbp]
\caption{Event yields in the $1999$--$2000$ and $2001$ datasets separated by tagging flavor and category.}
\smallskip
\begin{center}
\begin{tabular}{|c|ccc|ccc|ccc|} \hline
	 & \multicolumn{3}{|c|}{$1999$--$2000$} & \multicolumn{3}{|c|}{$2001$} & \multicolumn{3}{|c|}{Total}\\ \cline{2-10}
Category & $\Bz$  & $\Bzb$ & Tot & $\Bz$ & $\Bzb$ & Tot  & $\Bz$  & $\Bzb$ & Tot \rule[-1mm]{0mm}{+5mm} \\\hline\hline
Lepton   & $50$   & $59$   & $109$  & $25$  & $21$   & $46$   & $75$   & $80$	& $155$  \\[+1mm]
Kaon	 & $920$  & $877$  & $1797$ & $455$ & $468$  & $923$  & $1375$ & $1345$ & $2720$ \\[+1mm]
NT1	 & $215$  & $195$  & $410$  & $107$ & $92$   & $199$  & $322$  & $287$  & $609$  \\[+1mm]
NT2	 & $621$  & $560$  & $1181$ & $312$ & $236$  & $548$  & $933$  & $796$  & $1729$ \\[+1mm]
Untagged &  --    &  --    & $3103$ &	--  &	--   & $1425$ &  --    &  --	& $4528$ \\[+1mm]\hline
Total	 & $1806$ & $1691$ & $6600$ & $899$ & $817$  & $3141$ & $2705$ & $2508$ & $9741$ \rule[-2mm]{0mm}{6mm} \\\hline\hline
\end{tabular}
\end{center}
\label{tab:yields}
\end{table}

\subsection{Resolution function for $\deltat$}
\label{sec:dtres}
The time difference $\deltat$ is obtained from the measured distance between the
$z$ position of the $\Bhh$ and $\Btag$ vertices and the known boost of the CM 
frame.  The $z$ position of the \Btag\ vertex 
is determined with an iterative procedure that removes tracks with a large contribution
to the total $\chi^2$~\cite{BabarSin2betaObs,ref:babarmix}.  
An additional constraint is constructed from the three-momentum
and vertex position of the \Bhh\ candidate, and the average $\epem$ interaction point
and boost.  The typical $\deltaz$ resolution is $180\mum$.  
We require $\left|\deltat\right|<17\ps$ and $0.3 < \sigma_{\deltat} < 3.0\ps$, where 
$\sigma_{\deltat}$ is the event-by-event error on $\deltat$.  
The resolution function for 
signal candidates is identical to the one described in Ref.~\cite{BabarSin2betaObs}, with parameters 
determined from a fit to the combined tagged and untagged \Bflav\ sample.  The 
background resolution function is parameterized as the sum of three Gaussians, with the 
parameters determined from a fit to the on-resonance sideband data.  For both signal and 
background, the resolution function parameters are different for data collected in 
$1999$--$2000$ and $2001$ due to improved alignment of the SVT in more recent data.

\subsection{Background discrimination}
\label{sec:fisher}
The selected data sample contains $97\%$ background, mostly due to random
combinations of tracks produced in $\epem\to q\bar{q}$ events ($q = u,d,s,c$).
Discrimination of signal from background in the maximum likelihood fit is enhanced
by the use of a Fisher discriminant ${\cal F}$~\cite{twobodyPRL}.  The Fisher variables 
are constructed from the scalar sum of the CM momenta of all tracks and photons 
(excluding tracks from the \Bhh\ candidate) entering nine concentric cones centered 
on the thrust axis of the \Bhh\ candidate.  Background events dominantly contribute 
to the cones closest to the thrust axis, while the more spherical $\BB$ events distribute 
momentum more evenly.  
The distribution of ${\cal F}$ for signal events is parameterized 
as a single Gaussian, with parameters determined from Monte Carlo simulated decays and 
validated with data $\Bub\to\Dz\pim$ decays.  The background shape is parameterized
as the sum of two Gaussians, with parameters determined directly in the maximum
likelihood fit.

\subsection{Particle identification}
\label{sec:PID}
Identification of \Bhh\ tracks as pions or kaons is accomplished with the Cherenkov
angle measurement from the DIRC.  We construct Gaussian probability density functions (PDFs)
from the difference between measured and expected values of $\theta_c$ for the pion or kaon hypothesis, 
normalized by the resolution.  The DIRC performance is parameterized using a data sample of 
$D^{*+}\to\Dz\pip$, $\Dz\to \Km\pip$ decays.  Within the statistical precision of the control
sample, we find similar response for positive and negative tracks and use a single 
parameterization for both.  The performance of the DIRC has improved in the $2001$ dataset
due to a better aligned detector and improvements in the $\theta_c$ reconstruction 
algorithm.  We therefore use different parameter sets for the two running periods.

\subsection{Fit technique}
\label{sec:mlfit}
We use an unbinned extended maximum likelihood fit to extract yields and $\CP$ parameters
from the $\hh$ sample.  The likelihood for candidate $j$ tagged in category 
$c$ is obtained by summing the product of event yield $n_{i}$, tagging efficiency $\epsilon_{i,c}$,
and probability ${\cal P}_{i,c}$ over the $M$ possible signal and background hypotheses $i$,
\begin{equation}
{\cal L}_c = \exp{\left(-\sum_{i=1}^{M}n_i\epsilon_{i,c}\right)}
\prod_{j=1}^{N_c}\left[\sum_{i=1}^{M}n_i\epsilon_{i,c}{\cal P}_{i,c}(\vec{x}_j;\vec{\alpha}_i)\right].
\end{equation}
Due to low statistics in the $\pi\pi$ channel, we fix the tagging efficiencies $\epsilon_i$ and 
fit for the total yield in each component, rather than directly determining the yield in each
tagging category.  The probabilities ${\cal P}_{i,c}$ are evaluated as the product of probability
density functions (PDFs) for each of the independent variables 
$\vec{x}_j = \left\{\mes, \de, {\cal F}, \theta_c^+, \theta_c^-, \deltat\right\}$, where $\theta_c^+$
and $\theta_c^-$ are the Cherenkov angles for the positive and negative tracks, respectively.
The total likelihood is the product of likelihoods for each category
%\begin{equation}
%{\cal L} = \prod_{c=1}^{5}{\cal L}_c,
%\end{equation}
and the parameters are determined by minimizing the quantity $-2\ln{\cal L}$.

The $\deltat$ PDF for signal $\pip\pim$ decays is given by Eq.~\ref{fplusminus}, modified to
include the dilution and dilution difference for each tagging category, and convolved with 
the signal resolution function.  The $\deltat$ PDF for signal $K\pi$ events takes into account
$\Bz$--$\Bzb$ mixing,
depending on the charge of the kaon and the flavor of $\Btag$.  $\Bz\to\Kp\Km$ decays are 
parameterized as an exponential convolved with the resolution function.

There are $18$ free parameters in the fit:
\begin{itemize}
 \item $3$ signal and $3$ background yields ($n_i$) for the $\pi\pi$, $K\pi$, and $KK$
 hypotheses.
 \item Signal and background charge asymmetries (${\cal A}_{K\pi}$).
 \item $8$ background parameters describing the shapes in $\mes$, $\de$, and ${\cal F}$.
 \item $\spipi$ and $\cpipi$.
\end{itemize}
We fix the $B$ lifetime $\tau$ and mixing frequency $\deltamd$ to the PDG values~\cite{PDG}.

\section{Results}
\label{sec:Physics}
\begin{figure}[!tbp]
\begin{center}
\begin{minipage}[h]{8.0cm}
\includegraphics[width=8.0cm]{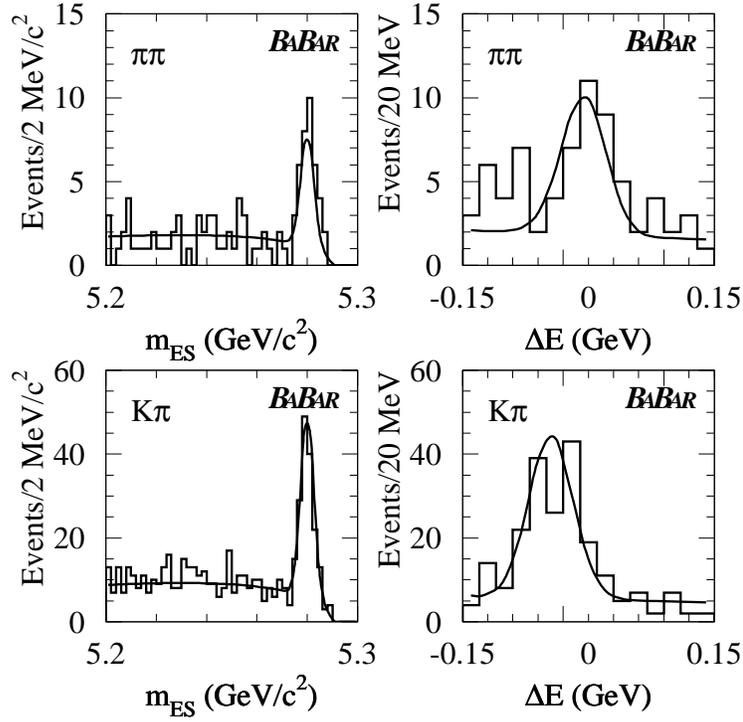}
\end{minipage}
\end{center}
\caption{Distributions of $\mes$ and $\de$ for events enhanced in signal 
$\pi^{\pm}\pi^{\mp}$ and $K^{\pm}\pi^{\mp}$ decays after likelihood ratio requirements.  The
solid curves represent projections of the maximum likelihood fit result after accounting
for the efficiency of the additional requirements.  The $\pi\pi\leftrightarrow K\pi$ cross-feed
is estimated to be less than three events in each plot.}
\label{fig:prplots}
\end{figure}

\begin{figure}[!tbp]
\begin{center}
\begin{minipage}[h]{6.0cm}
\includegraphics[width=6.0cm]{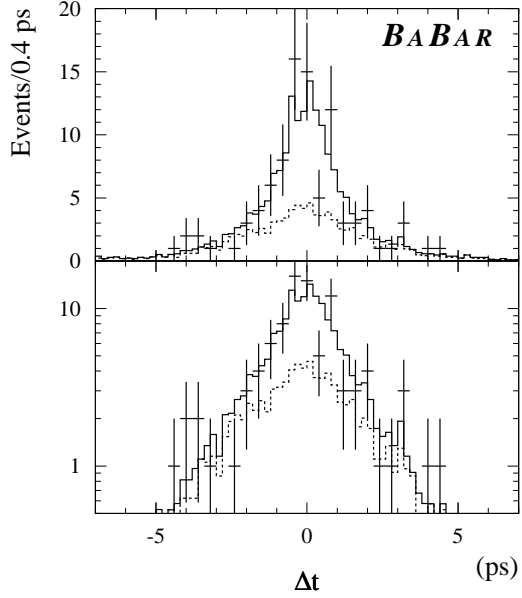}
\end{minipage}
\end{center}
\caption{Distribution of $\deltat$ for a sample enhanced in $\pi^{\pm}\pi^{\mp}$ events
obtained with likelihood ratio requirements.  The solid histogram represents the expected
distribution for signal and background, while the dashed histogram shows
the expected background shape.  The plot includes an estimated three $K\pi$ signal events.}
\label{fig:dtplot}
\end{figure}

In a sample of $33$ million $\BB$ pairs we find $65^{+12}_{-11}$ $\pi\pi$ and $217\pm 18$ $K\pi$ events
and measure the following $\CP$ parameters:
\begin{eqnarray*}
{\cal A}_{K\pi} & = & -0.07 \pm 0.08 \pm 0.02,\\
\spipi & = & 0.03^{+0.53}_{-0.56}\pm 0.11 ,\\
\cpipi & = & -0.25^{+0.45}_{-0.47}\pm 0.14,
\end{eqnarray*}
where the first error is statistical and the second is systematic.  The correlation between
$\spipi$ and $\cpipi$ is $-21\%$.  Figure~\ref{fig:prplots} shows distributions of $\mes$ and $\de$ 
for events enhanced in signal $\pi^{\pm}\pi^{\mp}$ and $K^{\pm}\pi^{\mp}$ decays based
on likelihood ratios.  The likelihood for a given signal or background hypothesis is constructed 
from the yield, tagging efficiency, and the product of probabilities for ${\cal F}$, $\theta_c$, 
and $\mes$\,($\de$) when plotting the projection for $\de$\,($\mes$).
The curves represent projections of the fit result
scaled by the efficiency of the additional requirements.  Figure~\ref{fig:dtplot} shows the 
$\deltat$ distribution for $\pipi$-enhanced events, with a looser selection than those applied 
in Fig.~\ref{fig:prplots}.  The solid histogram represents the expected distribution for the
selected sample, while the dashed histogram is the expected background shape.  The core is 
consistent with the estimated composition of $B$ decays and combinatorial background,
and the tails are described well by the background resolution function.

\subsection{Systematic uncertainties}
Several sources contribute to the systematic error on ${\cal A}_{K\pi}$, $\spipi$, and $\cpipi$:
\begin{itemize}
 \item {\bf PDFs for} {\boldmath $\mes$}, {\boldmath $\de$}, {\boldmath ${\cal F}$}.  We evaluate the
 systematic error on signal shapes with $\Bub\to\Dz\pim$ decays observed in data.  We
 obtain the background shapes directly in the fit and, in addition, use
 an asymmetric Gaussian as an alternative parameterization for ${\cal F}$.
 \item {\bf PDF for} {\boldmath $\theta_c$}.  We vary the PDF parameters within conservative ranges.
 \item {\bf Tagging}.  We vary efficiencies, dilutions, and dilution differences
 within their errors.  In addition, we compare tagging performance in simulated samples of 
 \Bflav\ and $\pip\pim$ decays, and repeat the maximum likelihhood fit with 
 the background tagging efficiencies as free parameters.
 \item {\bf PDFs for} {\boldmath $\deltat$}.  We vary all parameters of the signal and background
 resolution functions within their errors.  In addition, to account for possible
 effects due to SVT misalignment, we exchange the parameters for $1999$--$2000$
 and $2001$ data for both signal and background, which is a very conservative
 procedure.  We also compare the results of fits using parameters obtained
 separately from the tagged and untagged \Bflav\ samples.
 \item {\bf {\boldmath $\tau$} and {\boldmath $\deltamd$}}.  We vary these parameters within the 
 PDG errors~\cite{PDG}.
\end{itemize}
Table~\ref{tab:totalsys} summarizes the systematic errors coming from all 
sources, and the total systematic error calculated as the sum in quadrature of 
the individual uncertainties.

\begin{table}[!tbp]
\caption{Summary of systematic errors from all sources.  The total systematic error is calculated as the sum in
quadrature of the individual uncertainties.}
\begin{center}
\begin{tabular}{ccccccc} \hline\hline
				 & \multicolumn{2}{c}{${\cal A}_{K\pi}$} & \multicolumn{2}{c}{$S_{\pi\pi}$} & \multicolumn{2}{c}{$C_{\pi\pi}$} \\
 Source 		    & $+$	  & $-$ 	    &  $+$	  &  $-$	&  $+$        &  $-$	  \\ \hline
$\mes$  		    &  $0.003$  &  $0.002$  &  $0.007$  &  $0.005$    &  $0.018$  &  $0.022$\\
$\de$			    &  $0.014$  &  $0.013$  &  $0.009$  &  $0.035$    &  $0.096$  &  $0.110$\\
${\cal F}$		    &  $0.007$  &  $0.007$  &  $0.024$  &  $0.024$    &  $0.046$  &  $0.046$\\
$\theta_c$		    &  $0.004$  &  $0.004$  &  $0.021$  &  $0.022$    &  $0.038$  &  $0.041$\\
Sig Tagging		    &  $0.001$  &  $0.001$  &  $0.050$  &  $0.050$    &  $0.033$  &  $0.034$\\
Bkg Tagging		    &  $0.001$  &  $0.001$  &  $0.007$  &  $0.006$    &  $0.009$  &  $0.009$\\
Sig $\deltat$		    &  $0.001$  &  $0.001$  &  $0.068$  &  $0.069$    &  $0.032$  &  $0.027$\\
Bkg $\deltat$		    &  $0.002$  &  $0.002$  &  $0.052$  &  $0.053$    &  $0.020$  &  $0.020$\\
$\tau$ and $\deltamd$	    &  $0.000$  &  $0.000$  &  $0.011$  &  $0.011$    &  $0.007$  &  $0.007$\\
\hline
Total			    &  $0.017$  &  $0.016$  &  $0.106$  &  $0.111$  &  $0.125$  &  $0.136$  \rule[-4mm]{0mm}{10mm} \\
\hline \hline
\end{tabular}
\end{center}
\label{tab:totalsys}
\end{table}

\subsection{Validation}
\label{sec:validation}
Extensive studies using ``toy'' Monte Carlo, \geant\-3 Monte Carlo simulation, and data samples
have been used to validate the fit technique.  In large samples of toy Monte Carlo 
experiments generated with the statistics observed in the full dataset, we find no evidence of bias 
in any of the free parameters and the errors are consistent with expectations.  The most probable
errors are $0.59$ and $0.41$ for $\spipi$ and $\cpipi$, respectively, consistent with the 
data fit results.

Fitting large samples of pure $\pi\pi$ and $K\pi$ simulated Monte Carlo events, we are able
to extract the input values without bias when floating $\tau$, $\deltamd$, $\spipi$, and $\cpipi$.
In samples of simulated signal and background Monte Carlo events equivalent $10\invfb$ we obtain
consistent values of yields, $\tau$, $\deltamd$, and the $\CP$ parameters, where the
errors on the latter are in agreement with toy Monte Carlo predictions.

Fits to signal and background yields in the $1999$--$2000$ and $2001$ datasets
without $\deltat$ information give results consistent with our branching fraction 
measurement~\cite{twobodyPRL}, and we find consistent values of all fitted background 
parameters between the two datasets.  Addition of $\deltat$ in the likelihood function 
improves the statistical error on $N_{\pi\pi}$ by approximately $9\%$, while the yield 
changes by only $1$ event ($1.5\%$).

As a validation of the $\deltat$ parameterization in data, we fit the full dataset to
simultaneously extract yields, background parameters, $\tau$, $\deltamd$, $\spipi$, and 
$\cpipi$.  We find $\tau = (1.52\pm 0.12)\ps$ and $\deltamd = (0.54\pm 0.09)\hbar \ps^{-1}$, 
and the remaining free parameters are stable with respect to the fit with fixed $\tau$ and 
$\deltamd$.

\section{Summary}
\label{sec:Summary}
We have presented a preliminary measurement of the time-dependent $\CP$-violating
asymmetry in $B\to\pip\pim$ decays, and a preliminary updated measurement of 
the asymmetry between $\Bz\to\Kp\pim$ and $\Bzb\to\Km\pip$ decays.
In a sample of $33$ million $\BB$ pairs we observe $65^{+12}_{-11}$ $\pi\pi$ and 
$217\pm 18$ $K\pi$ candidates and measure the following parameters:
\begin{eqnarray*}
{\cal A}_{K\pi} & = & -0.07 \pm 0.08 \pm 0.02,\\
\spipi & = & 0.03^{+0.53}_{-0.56}\pm 0.11 ,\\
\cpipi & = & -0.25^{+0.45}_{-0.47}\pm 0.14,
\end{eqnarray*}
where the first error is statistical and the second is systematic.  The
systematic error on ${\cal A}_{K\pi}$ includes an uncertainty of $\pm 0.01$ from
possible charge bias in track reconstruction and particle identification.  We 
observe no evidence for direct \CP\ violation in $B\to K^{\pm}\pi^{\mp}$ decays, and calculate a 
$90\%$ confidence limit on ${\cal A}_{K\pi}$ of $\left[-0.21, +0.07\right]$ assuming 
Gaussian errors.  With the addition of more data and improvements in detector 
performance, measurements of ${\cal A}_{K\pi}$, $\spipi$, and $\cpipi$ will yield
increasingly more important information about \CP\ violation in the $B$ meson system.

\section{Acknowledgments}
\label{sec:Acknowledgments}

% Standard acknowledgments paragraph; must always be included.
\input pubboard/acknowledgements

%%%%%%%%%%%%%%%%%%      BIBLIOGRAPHY      %%%%%%%%%%%%%%%%%

%%%%%%%%%%%%%%%%%%%%%%%%%%%%%%%%%%%%%%%%%%%%%%%%%%%%%%%%%%%

\end{document}

%% file: pubboard/authors_EPS2001.tex
\begin{center}
\small

The \babar\ Collaboration,
\bigskip

%% author list as of 12-Jul-2001 (622 authors)
B.~Aubert,
D.~Boutigny,
J.-M.~Gaillard,
A.~Hicheur,
%A.~Jeremie, per J.P.Lees
Y.~Karyotakis,
J.~P.~Lees,
P.~Robbe,
V.~Tisserand
\inst{Laboratoire de Physique des Particules, F-74941 Annecy-le-Vieux, France }
A.~Palano
\inst{Universit\`a di Bari, Dipartimento di Fisica and INFN, I-70126 Bari, Italy }
G.~P.~Chen,
J.~C.~Chen,
N.~D.~Qi,
G.~Rong,
P.~Wang,
Y.~S.~Zhu
\inst{Institute of High Energy Physics, Beijing 100039, China }
% ...

\end{center}\newpage

%% file: pubboard/acknowledgements.tex
We are grateful for the 
extraordinary contributions of our \pep2\ colleagues in
achieving the excellent luminosity and machine conditions
that have made this work possible.
The collaborating institutions wish to thank 
SLAC for its support and the kind hospitality extended to them. 
This work is supported by the
US Department of Energy
and National Science Foundation, the
Natural Sciences and Engineering Research Council (Canada),
Institute of High Energy Physics (China), the
Commissariat \`a l'Energie Atomique and
Institut National de Physique Nucl\'eaire et de Physique des Particules
(France), the
Bundesministerium f\"ur Bildung und Forschung
(Germany), the
Istituto Nazionale di Fisica Nucleare (Italy),
the Research Council of Norway, the
Ministry of Science and Technology of the Russian Federation, and the
Particle Physics and Astronomy Research Council (United Kingdom). 
Individuals have received support from the Swiss 
National Science Foundation, the A. P. Sloan Foundation, 
the Research Corporation,
and the Alexander von Humboldt Foundation.